\documentclass[%
 aip,
rsi,
 amsmath,amssymb,
 reprint,%
]{revtex4-1}

\usepackage{graphicx}

\usepackage{hyperref}
\hypersetup{
    colorlinks=true,
    linkcolor=blue,
    citecolor=blue,
    urlcolor=blue
}

\begin{document}


\title{A dual-region ytterbium tweezer apparatus with an in-vacuum nanofiber cavity} 

\author{Shunichiro Hashimoto}
\affiliation{Nanofiber Quantum Technologies, Inc., 1-22-3 Nishiwaseda, Shinjuku-ku, Tokyo 169-0051, Japan}
\affiliation{Department of Applied Physics, Waseda University, 3-4-1 Okubo, Shinjuku-ku, Tokyo 169-8555, Japan}

\author{Hideki Ozawa}
\affiliation{Nanofiber Quantum Technologies, Inc., 1-22-3 Nishiwaseda, Shinjuku-ku, Tokyo 169-0051, Japan}

\author{Yusuke Hisai}
\affiliation{Nanofiber Quantum Technologies, Inc., 1-22-3 Nishiwaseda, Shinjuku-ku, Tokyo 169-0051, Japan}

\author{Takao Aoki}
\affiliation{Nanofiber Quantum Technologies, Inc., 1-22-3 Nishiwaseda, Shinjuku-ku, Tokyo 169-0051, Japan}
\affiliation{Department of Applied Physics, Waseda University, 3-4-1 Okubo, Shinjuku-ku, Tokyo 169-8555, Japan}
\affiliation{RIKEN Center for Quantum Computing (RQC), RIKEN, 2-1 Hirosawa, Wako, Saitama, 351-0198, Japan.}

\author{Ryotaro Inoue}
\email{ryotaro.inoue@nano-qt.com}
\affiliation{Nanofiber Quantum Technologies, Inc., 1-22-3 Nishiwaseda, Shinjuku-ku, Tokyo 169-0051, Japan}

\date{22 September 2026}
\begin{abstract}
We present a dual-region experimental apparatus for trapping single atoms of the fermionic isotope ytterbium-171 (${}^{171}\mathrm{Yb}$) with optical tweezers in a science chamber containing an in-vacuum nanofiber cavity. 
A source region is located approximately $240\,\mathrm{mm}$ from the science chamber, preserving optical access around the chamber for high-numerical-aperture tweezer optics. 
A pulsed transport sequence based on longitudinal moving molasses launches atomic pulses toward the science chamber, where they are decelerated and recaptured in a magneto-optical trap.
We subsequently load single atoms into a $2\times2$ tweezer array located approximately $50\,\mathrm{\mu m}$ from the nanofiber and measure a $1/e$ lifetime of $9.5(7)\,\mathrm{s}$. 
These results demonstrate that the dual-region architecture provides sufficient ${}^{171}\mathrm{Yb}$ atoms for single-atom tweezer operation near an in-vacuum nanofiber cavity.
\end{abstract}


\maketitle 

\section{Introduction}

Optical tweezer arrays provide a versatile platform for controlling individual neutral atoms and have enabled rapid progress in neutral-atom-based quantum information processing \cite{pichardRearrangementIndividualAtoms2024, manetschTweezerArray61002025, everedHighfidelityParallelEntangling2023, bluvsteinLogicalQuantumProcessor2024}. Integrating such arrays with optical interfaces is an important route toward connecting local atomic processors and realizing distributed atom-photon architectures \cite{monroeLargescaleModularQuantumcomputer2014, coveyQuantumNetworksNeutral2023, sunamiScalableNetworkingNeutralAtom2025,Takahata2026}. 
In particular, incorporating in-vacuum photonic devices, including optical nanofiber cavities \cite{horikawaLowlossTelecombandNanofiber2025}, into a tweezer apparatus imposes additional constraints on the science-chamber design.

For this type of apparatus, the total number of cold atoms in the science chamber is not necessarily the sole relevant figure of merit. 
Efficient tweezer loading requires a sufficient local atom number or density, while excessive exposure of an in-vacuum photonic device to untrapped atoms should be avoided because adsorption on its surfaces can alter its properties or performance \cite{laiTransmissionDegradationPreservation2013, andersonOpticalNanofiberTemperature2018}.
Spatially separating the atom source from the science chamber enables the delivered atomic flux to be tailored to the requirements of downstream loading, thereby addressing these competing requirements while also preserving optical access around the science chamber for the tweezer system.
More generally, similar spatial separation of the atom-loading and science regions has been employed in recent neutral-atom architectures \cite{norciaIterativeAssembly1712024a,gygerContinuousOperationLargescale2024,chiuContinuousOperationCoherent2025}, in which optically trapped atomic reservoirs are transported from the loading region to the science region repeatedly or continuously, depending on the operating scheme.
This arrangement allows fresh atoms to be prepared and supplied with minimal disruption to operations on atoms already held in the science region.

\begin{figure*}[tb]
    \centering
    \includegraphics[width=\textwidth]{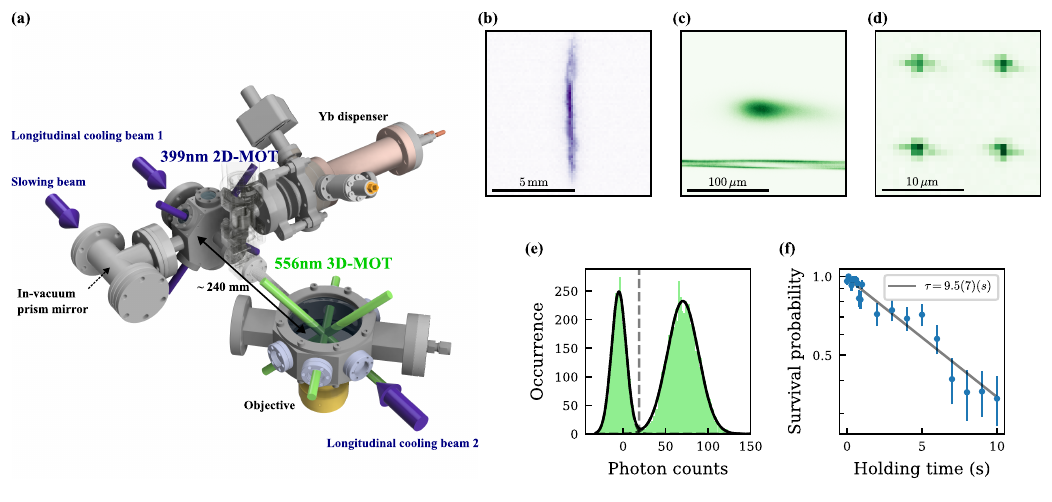}
    \caption{Overview of the dual-region ${}^{171}\mathrm{Yb}$ apparatus and single-atom tweezer operation in the science chamber. (a) Schematic of the apparatus. The source region contains the Yb oven and a $399\,\mathrm{nm}$ 2D-MOT. Cold Yb atoms are transported over approximately $240\,\mathrm{mm}$ to the spatially separated science chamber using a pulsed sequence based on longitudinal moving molasses, decelerated near the science chamber by the $399\,\mathrm{nm}$ beams, and subsequently recaptured in a $556\,\mathrm{nm}$ 3D-MOT. The science chamber accommodates the optical-tweezer system and an in-vacuum nanofiber cavity. (b) Fluorescence image of Yb atoms in the $399\,\mathrm{nm}$ 2D-MOT with no longitudinal cooling applied. (c) Fluorescence image of Yb atoms in the $556\,\mathrm{nm}$ 3D-MOT. The horizontal line visible below the atomic cloud is $556\,\mathrm{nm}$ cooling light scattered by the nanofiber in the science chamber. (d) An averaged fluorescence image of single atoms trapped in a $2\times2$ optical-tweezer array in the science chamber, approximately $50\,\mathrm{\mu m}$ from the nanofiber. (e) Photon-count histogram summed over all tweezer sites. The dashed line indicates the threshold used to distinguish empty and occupied sites. The observed single-atom filling fraction is $66.4(5)\%$ for the specific loading conditions used here. With further optimization of the loading conditions in the same system, a filling fraction of $74.9(7)\%$ was obtained \cite{Ozawa2026}.
    (f) Survival probability as a function of the hold time in the optical-tweezer array. An exponential fit gives a $1/e$ trap lifetime of $9.5(7)\,\mathrm{s}$.
    }
    \label{fig:apparatus}
\end{figure*}

In this work, we describe a dual-region ytterbium-171 (${}^{171}\mathrm{Yb}$) apparatus in which the source region is spatially separated from a science chamber containing an in-vacuum nanofiber cavity. 
The apparatus has also been used to demonstrate atom--photon coupling mediated by the nanofiber cavity~\cite{Ozawa2026}.
Atoms are collected and cooled in a permanent-magnet $399\,\mathrm{nm}$ two-dimensional magneto-optical trap (2D-MOT), transported to the science chamber using a pulsed sequence based on $399\,\mathrm{nm}$ moving molasses, and recaptured in a $556\,\mathrm{nm}$ three-dimensional MOT (3D-MOT).
The fermionic isotope ${}^{171}\mathrm{Yb}$ is particularly attractive for neutral-atom quantum technologies because its nuclear spin-$1/2$ degree of freedom is available in both the ${}^{1}\mathrm{S}_{0}$ ground state and the long-lived ${}^{3}\mathrm{P}_{0}$ metastable state \cite{jenkinsYtterbiumNuclearSpinQubits2022, maUniversalGateOperations2022, nakamuraHybridAtomTweezer2024}. 
Although the initial cooling and deceleration of ytterbium are commonly performed using the strong $399\,\mathrm{nm}$ ${}^{1}\mathrm{S}_{0}$--${}^{1}\mathrm{P}_{1}$ transition, generating high optical powers at this wavelength can add substantial complexity to the optical system.
Established approaches include frequency doubling of high-power $798\,\mathrm{nm}$ sources \cite{pizzocaroEfficientFrequencyDoubling2014} and injection locking of blue laser diodes \cite{hosoyaInjectionLockingHigh2015}.
This motivates the development of efficient atom-loading and transport schemes that operate with a modest $399\,\mathrm{nm}$ optical-power budget.
In this work, we characterize the dependence of the delivered atom number on the source and transport parameters and identify an operating regime that provides sufficient atoms for downstream tweezer loading.
Finally, we demonstrate single-atom tweezer loading in the science chamber with the nanofiber cavity installed.

\section{Experimental apparatus}

Figure~\ref{fig:apparatus} shows an overview of the apparatus. 
The source region contains an effusive Yb dispenser and a 2D-MOT. 
It is connected to a separate science chamber through a gate valve and a flexible tube.
The distance between the 2D-MOT region and the science chamber, which accommodates an in-vacuum nanofiber cavity, is approximately $240\,\mathrm{mm}$.  

The science chamber is evacuated by an ion pump (VacIon Plus 40 StarCell, Agilent Technologies) together with a non-evaporable getter pump (CapaciTorr Z200, SAES Getters). With the atom source operating and the nanofiber cavity in the science chamber, the pressure indicated by the ion pump remains below $1\times10^{-8}\,\mathrm{Pa}$, with the ion current below the controller's detection threshold.  We characterize the integrated performance of the science chamber through single-atom lifetime measurements in Sec.~\ref{sec:tweezer}. Optical tweezers are formed using $759\,\mathrm{nm}$ light focused by a high-numerical-aperture objective lens. 
In the measurements reported here, a $2\times2$ tweezer array is generated approximately $50\,\mathrm{\mu m}$ from the nanofiber.

\subsection{Source region}

The Yb oven consists of a Yb dispenser mounted on an electrical feedthrough, following a design similar to that reported in Ref.~\cite{nomuraDirectLoadingYb2023}. The dispenser is surrounded by an ICF70 nipple made of 0.2 wt\% Be-Cu alloy instead of the usual stainless steel. 
The low emissivity of the material reduces the absorption of thermal radiation from the hot dispenser, while its high thermal conductivity efficiently dissipates the absorbed heat, limiting the temperature rise of the vacuum wall and consequently reducing thermally induced outgassing. 
Hot Yb atoms emitted from the dispenser pass through an additional radiation shield and a differential-pumping tube with a diameter of $4\,\mathrm{mm}$ and a length of $40\,\mathrm{mm}$ before entering the 2D-MOT region. 
The pressure on the oven side is maintained at $3 \times 10^{-7}\,\mathrm{Pa}$ during the oven operation by a $2\,\mathrm{L/s}$ ion pump (VacIon 2L/s Pump, Agilent Technologies), while the 2D-MOT region is pumped by a combination pump (NexTorr Z200, SAES Getters). Initial cooling and slowing of the emitted Yb atoms use the dipole-allowed $399\,\mathrm{nm}$ ${}^{1}\mathrm{S}_{0}$--${}^{1}\mathrm{P}_{1}$ transition with linewidth $\Gamma_{399}/2\pi=29\,\mathrm{MHz}$. 
A counter-propagating slowing beam is introduced through an in-vacuum prism mirror. The transverse 2D-MOT beam is a retroreflected Gaussian beam with a waist radius of $4.7\,\mathrm{mm}$. 
Here, the magnetic-field gradient for the 2D-MOT operation is produced by permanent magnets arranged similarly to those in Refs.~\cite{lamporesiCompactHighfluxSource2013, nosskeTwodimensionalMagnetoopticalTrap2017, hosoyaHighfluxColdYtterbium2023}. 
We use samarium-cobalt magnets for their relatively low temperature coefficient, which improves magnetic field stability under temperature variations. 
The relevant magnetic-field gradient is approximately $100\,\mathrm{G/cm}$.
The optical powers available for the 2D-MOT characterization are $20\,\mathrm{mW}$ for the transverse cooling beam and $15\,\mathrm{mW}$ for the slowing beam. As shown in Sec.~\ref{sec:delivery_characterization}, these powers are sufficient to provide the atom number required for the downstream 3D-MOT and tweezer-loading sequence.

\subsection{Atom transport to the 3D-MOT}

Two counter-propagating, linearly cross-polarized $399\,\mathrm{nm}$ beams (a $\mathrm{lin}\perp\mathrm{lin}$ configuration) are applied along the transport axis.
During loading of the 2D-MOT, their detuning is set to $-1.4\Gamma_{399}$, providing longitudinal cooling in the laboratory frame. 
To launch the atoms toward the science chamber, the beam frequencies are swept to shift the moving molasses frame, then held near the final frequencies for a short period. 
A key consideration in designing the transport sequence is suppressing transverse heating and the resulting spatial expansion caused by photon scattering. 
During the initial launch, transverse cooling by the 2D-MOT is available only within the source region. 
The launch is therefore completed while the atoms remain within, or close to, this transversely cooled region. 
After the longitudinal beams are turned off, the atoms propagate ballistically over most of the distance toward the science chamber, thereby minimizing unnecessary photon scattering outside the 2D-MOT cooling region.

Near the science chamber, the longitudinal $399\,\mathrm{nm}$ beams are reapplied at their original laboratory-frame detuning of $-1.4\Gamma_{399}$ for $2\,\mathrm{ms}$ to decelerate the atomic pulse. 
Excessive deceleration is undesirable because the associated photon scattering produces additional transverse heating, while the resulting lower longitudinal velocity increases the time available for transverse expansion and gravitational displacement. 
The deceleration is therefore limited to bringing the atoms into the velocity-capture range of the downstream 3D-MOT. 
Its timing is chosen so that deceleration begins as close as possible to the 3D-MOT region while still allowing the spatially extended atomic pulse to be sufficiently slowed before it reaches the edge of the capture region.
Following the $399\,\mathrm{nm}$ deceleration stage, the 3D-MOT is turned on for $30\,\mathrm{ms}$ to capture and cool the transported atoms on the intercombination $556\,\mathrm{nm}$ ${}^{1}\mathrm{S}_{0}$--${}^{3}\mathrm{P}_{1}$ transition with linewidth $\Gamma_{556}/2\pi=182\,\mathrm{kHz}$. 

The frequency sweep during the launch changes the relative detunings of the two $399\,\mathrm{nm}$ beams, thereby accelerating the moving-molasses frame and the atoms following it toward the science chamber.
The full sweep corresponds to frequency shifts of approximately $\pm50\,\mathrm{MHz}$, giving laboratory-frame detunings of approximately $-3.1\Gamma_{399}$ and $0.3\Gamma_{399}$ for the two beams. 
These values correspond to a moving-molasses velocity of approximately $20\,\mathrm{m/s}$.
The actual launch velocity depends on the launch (acceleration) duration $t_\text{launch}$ and the temporal profile of the launch sequence; the launch conditions are therefore optimized experimentally.
Since the photon recoil velocity on the $399\,\mathrm{nm}$ transition is $v_R\simeq6\,\mathrm{mm/s}$, a change in atomic velocity of $20\,\mathrm{m/s}$ corresponds to several thousand photon recoils.
The broad linewidth of the $399\,\mathrm{nm}$ transition enables both acceleration and deceleration on a millisecond timescale.

\begin{figure}[tbp]
    \centering
    \includegraphics[width=\linewidth]{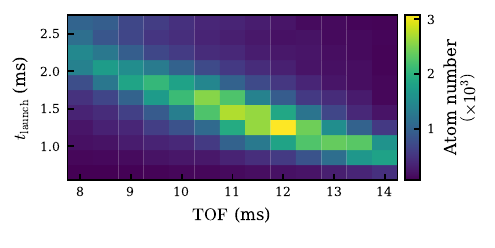}
    \caption{Characterization of the pulsed transport sequence. The number of atoms recaptured in the 3D-MOT is shown as a function of time of flight (TOF) and acceleration duration ($t_\text{launch}$). The duration of the downstream $399\,\mathrm{nm}$ deceleration pulse is fixed at $2\,\mathrm{ms}$.
    Each point represents the average of five experimental realizations.
    }
    \label{fig:transport_optimization}
\end{figure}

\begin{figure*}[tb]
    \centering
    \includegraphics[width=\linewidth]{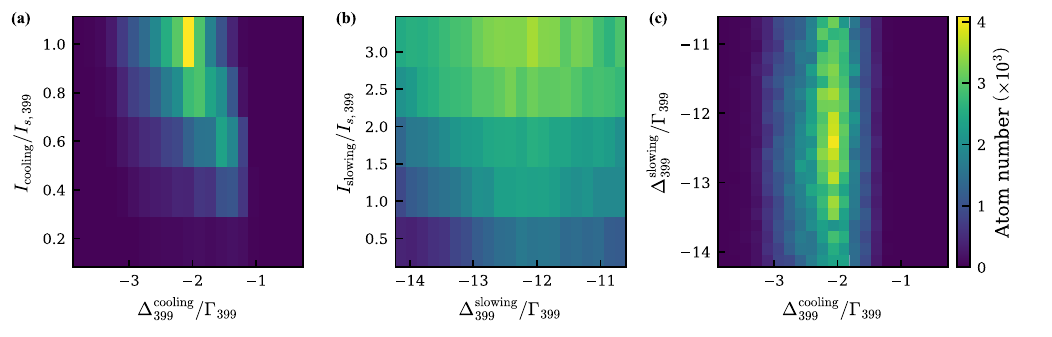}
    \caption{Characterization of the 2D-MOT source. The number of atoms delivered by a single transport pulse and recaptured in the 3D-MOT is shown as a function of the detuning and intensity of (a) the transverse 2D-MOT cooling beam and (b) the slowing beam. (c) Recaptured atom number as a function of the cooling- and slowing-beam detunings at $(I_{\mathrm{cooling}},I_{\mathrm{slowing}})\simeq(1.0,3.1)I_{\mathrm{s},399}$. 
    Within the available power range, the atom number increases with beam intensity and no power optimum is observed. At the operating detunings $(\Delta^{\mathrm{cooling}}_{399},\Delta^{\mathrm{slowing}}_{399})\simeq(-2.1,-12.5)\Gamma_{399}$ and the maximum available cooling- and slowing-beam powers, a single transport pulse delivers up to $4.0\times10^{3}$ atoms to the 3D-MOT. 
    In (a) and (b), each point represents the average of five experimental realizations, whereas (c) shows single-shot data.
    }
    \label{fig:2D_MOT_characterization}
\end{figure*}

\section{Characterization of atom delivery}
\label{sec:delivery_characterization}

The source and transport system is designed to deliver a sufficient number of atoms from the spatially separated source region to the downstream 3D-MOT and tweezer-loading sequence. 
Fluorescence from both atomic ensembles and tweezer-trapped atoms is recorded with the same camera (C15550-20UP, Hamamatsu Photonics). 
For the ensemble measurements, the number of atoms is determined from the detected fluorescence, accounting for the photon-collection numerical aperture, camera quantum efficiency, and the optical transmission from the atoms to the camera.

\subsection{Transport characteristics}

During the transport measurements, the $556\,\mathrm{nm}$ 3D-MOT cooling-beam intensity and detuning are fixed at $14\,I_{\mathrm{s},556}$ and $-0.97\,\Gamma_{556}$, respectively. The two longitudinal $399\,\mathrm{nm}$ beams each have a peak intensity of $0.1\,I_{\mathrm{s},399}$. Here, $I_{\mathrm{s},556}$ and $I_{\mathrm{s},399}$ denote the saturation intensities of the corresponding transitions.

Figure~\ref{fig:transport_optimization} shows the number of atoms recaptured in the 3D-MOT as a function of the total acceleration duration $t_\text{launch}$ and the subsequent time of flight (TOF). Here, $t_\text{launch}$ includes both the frequency sweep and the hold at the target detuning. 
The largest recaptured atom number is obtained around $t_\text{launch}\simeq1.2\,\mathrm{ms}$ and $\mathrm{TOF}\simeq12\,\mathrm{ms}$. 
The optimum TOF decreases systematically with increasing $t_\text{launch}$, consistent with the larger forward velocity resulting from a longer launch duration.
Consequently, a shorter ballistic propagation time is required to deliver the atoms to the downstream deceleration and 3D-MOT region at the appropriate time. 
The broad optimum observed in Fig.~\ref{fig:transport_optimization} is consistent with the transport strategy described above.

\subsection{2D-MOT loading characteristics}

Figure~\ref{fig:2D_MOT_characterization}(a) and (b) show the atom number recaptured in the 3D-MOT as functions of the detuning and intensity of the transverse 2D-MOT cooling beam and the slowing beam, respectively. 
The accessible intensity range is set by the available optical powers, $20\,\mathrm{mW}$ for the transverse cooling beam and $15\,\mathrm{mW}$ for the slowing beam. 
Within this range, the delivered atom number increases monotonically with increasing cooling- and slowing-beam intensities. 
The largest available powers are used for the measurements in Fig.~\ref{fig:2D_MOT_characterization}(c). 
The optimum with respect to optical power is therefore not reached within the available range; nevertheless, the maximum available powers already provide an atom number sufficient for the downstream tweezer-loading sequence.

The two-dimensional detuning scan in Fig.~\ref{fig:2D_MOT_characterization}(c) identifies an operating region around $(\Delta^{\mathrm{cooling}}_{399},\Delta^{\mathrm{slowing}}_{399})\simeq(-2.1,-12.5)\Gamma_{399}$. Under these conditions, one transport pulse provides up to $4.0\times10^{3}$ atoms in the 3D-MOT. 
This number is sufficient for the subsequent single-atom tweezer-loading procedure. 
Additional loading and transport cycles can be applied when a larger accumulated number of atoms is required.

\subsection{Single-atom loading in the science chamber}
\label{sec:tweezer}
To demonstrate the operation of the complete dual-region apparatus, we load single atoms into optical tweezers in the science chamber with the nanofiber cavity installed.
We generate optical-tweezer arrays with a spatial light modulator (X15213-02, Hamamatsu Photonics) using $759\,\mathrm{nm}$ light \cite{KohnoOneDimensional2009,lemkeSpin122009}.
The trap depth is approximately $340\,\mathrm{\mu K}$. For the measurements reported here, a $2\times2$ array is generated in the science chamber, approximately $50\,\mathrm{\mu m}$ from the nanofiber.

Following the recapture of the transported atoms in the $556\,\mathrm{nm}$ 3D-MOT, the cooling-light intensity and detuning, together with the magnetic-field gradient, are adjusted to load atoms into the tweezers, which can initially be multiply occupied.
After the quadrupole field is switched off and a bias field is applied, the cooling-light parameters are adjusted for further cooling and to induce light-assisted collisions \cite{schlosserCollisionalBlockadeMicroscopic2002, grunzweigNeardeterministicPreparationSingle2010, jenkinsYtterbiumNuclearSpinQubits2022}.
Under the specific loading conditions used for the measurements reported here, the observed single-atom filling fraction is $66.4(5)\%$, as illustrated by the fluorescence image and photon-count histogram in Fig.~\ref{fig:apparatus}(d) and (e). 
With further optimization of the loading conditions in the same system, a filling fraction of $74.9(7)\%$ was obtained \cite{Ozawa2026}.

As an integrated performance test of the apparatus with the nanofiber cavity and its associated in-vacuum hardware installed, we measure the survival probability of individually trapped atoms as a function of the hold time.
As shown in Fig.~\ref{fig:apparatus}(f), an exponential fit gives a $1/e$ trap lifetime of $9.5(7)\,\mathrm{s}$, demonstrating multi-second single-atom trapping in the complete science chamber configuration.

A comparable lifetime measurement performed in the same system under different operating conditions yielded $6.8(4)\,\mathrm{s}$. Both measurements demonstrate multi-second single-atom trapping under representative operating conditions.
In addition to the single-atom trapping performance, we also examine the intrinsic finesse \cite{horikawaLowlossTelecombandNanofiber2025} of the nanofiber cavity over several months and, in some cases, longer periods of routine atom-loading operation.
The apparatus design described here has also been implemented in a second experimental system.
Across the two systems, nanofiber cavities with an intrinsic finesse of approximately $10^3$ upon installation have shown no significant decrease from their initial values over experimentally relevant timescales.

\section{Conclusion}

We have developed and characterized a dual-region ${}^{171}\mathrm{Yb}$ cold-atom apparatus for optical-tweezer experiments in a science chamber containing an in-vacuum nanofiber cavity. 
The architecture spatially separates the source region from the science chamber by approximately $240\,\mathrm{mm}$, preserving optical access around the science chamber. 
Cold atoms are transported from the source region to the science chamber using a pulsed sequence based on longitudinal moving molasses operating on the $399\,\mathrm{nm}$ transition. 
Using $20\,\mathrm{mW}$ of transverse cooling-beam power and $15\,\mathrm{mW}$ of slowing-beam power at $399\,\mathrm{nm}$, the system delivers up to $4.0\times10^{3}$ atoms to the $556\,\mathrm{nm}$ 3D-MOT. 
Under these conditions, we further demonstrate single-atom loading into a $2\times2$ tweezer array located approximately $50\,\mathrm{\mu m}$ from the nanofiber and measure a $1/e$ trap lifetime of $9.5(7)\,\mathrm{s}$. These results establish the viability of a dual-region architecture that spatially separates atom preparation from a science chamber incorporating both an optical-tweezer system and an in-vacuum nanofiber cavity.

\begin{acknowledgments}
We thank the members of the NanoQT team for their assistance. 
This work was supported by JST Moonshot R\&D (Grant Number JPMJMS2268) and JST SPRING (Grant Number JPMJSP2128). This paper is based on results obtained from a project,
JPNP20017, subsidized by the New Energy and Industrial Technology
Development Organization (NEDO).

\end{acknowledgments}

\section*{author declarations}
\subsection*{Conflict of Interest}
S.H., H.O., Y.H., and R.I.~are employees of Nanofiber Quantum Technologies, Inc.
T.A.~is a co-founder of and an equity shareholder in Nanofiber Quantum Technologies, Inc.

\subsection*{Author Contributions}
\noindent
\textbf{Shunichiro Hashimoto}: 
Data curation (lead); Formal analysis (equal); Investigation (equal); Software (equal); Validation (equal); Visualization (lead); Writing – original draft (supporting).
\textbf{Hideki Ozawa}: 
Conceptualization (supporting); Formal analysis (equal); Investigation (equal); Methodology (equal); Software (equal); Writing – review \& editing (supporting).
\textbf{Yusuke Hisai}: Investigation (equal); Writing – review \& editing (supporting).
\textbf{Takao Aoki}: 
Funding acquisition (lead); Project administration (supporting); Writing – review \& editing (supporting).
\textbf{Ryotaro Inoue}: 
Conceptualization (lead); Formal analysis (equal); Methodology (equal); Project administration (lead); Supervision (lead); Validation (equal); Writing – original draft (lead); Writing – review \& editing (lead).

\section*{data availability}
The data that support the findings of this study are available from the corresponding author upon reasonable request.

\section*{References}
\bibliography{reference}

\end{document}